\documentclass[twoside,twocolumn,9pt]{article}
\usepackage{extsizes}
\usepackage[super,sort&compress,comma]{natbib} 
\usepackage[left=1.5cm, right=1.5cm, top=1.785cm, bottom=2.0cm]{geometry}
\usepackage{balance}

\usepackage{sectsty}
\allsectionsfont{\scshape}

\usepackage{graphicx} 
\usepackage{lastpage}
\usepackage{float}
\usepackage{fancyhdr}
\usepackage{fnpos}
\usepackage[english]{babel}
\usepackage{array}
\usepackage[T1]{fontenc}
\usepackage[usenames,dvipsnames]{xcolor}
\usepackage{setspace}
\usepackage{hyperref}
\usepackage{subfigure}

\makeatletter
\def\blfootnote{\xdef\@thefnmark{}\@footnotetext}
\makeatother

\usepackage{upgreek}
\usepackage{xspace}
\usepackage{placeins}
\usepackage{amsmath}
\usepackage{amssymb}

\usepackage{mciteplus}
\usepackage{adjustbox}
\usepackage{epstopdf}
\usepackage{stfloats}

\begin{document}

\twocolumn[
  \begin{@twocolumnfalse}
\vspace{3cm}

\begin{center}

    \noindent\huge{\textbf{\textsc{Deposition and alignment of fiber suspensions by dip coating}}} \\
    \vspace{1cm}

    \noindent\large{Deok-Hoon Jeong,\textit{$^{a}$} Langqi Xing,\textit{$^{a}$} Michael Ka Ho Lee,\textit{$^{a}$} Nathan Vani,\textit{$^{a,b}$} Alban Sauret \textit{$^{a}$}}$^{\ast}$ \\

    \vspace{5mm}
   \noindent\large{\today} \\

    \vspace{1cm}
    \textbf{\textsc{Abstract}}
    \vspace{2mm}

\end{center}

\noindent\normalsize{
   The dip coating of suspensions made of monodisperse non-Brownian spherical particles dispersed in a Newtonian fluid leads to different coating regimes depending on the ratio of the particle diameter to the thickness of the film entrained on the substrate. In particular, dilute particles dispersed in the liquid are entrained only above a threshold value of film thickness. In the case of anisotropic particles, in particular fibers, the smallest characteristic dimension will control the entrainment of the particle. Furthermore, it is possible to control the orientation of the anisotropic particles depending on the substrate geometry. In the thick film regime, the Landau-Levich-Derjaguin remains valid if one account for the change in viscosity. To test the hypotheses, we performed dip-coating experiments with dilute suspensions of non-Brownian fibers with different length-to-diameter aspect ratios. We characterize the number of fibers entrained on the surface of the substrate as a function of the withdrawal velocity, allowing us to estimate a threshold capillary number below which all the particles remain in the liquid bath. Besides, we measure the angular distribution of the entrained fibers for two different substrate geometries: flat plates and cylindrical rods. We then measure the film thickness for more concentrated fiber suspensions. The entrainment of the fibers on a flat plate and a cylindrical rod is primarily controlled by the smaller characteristic length of the fibers: their diameter. At first order, the entrainment threshold scales similarly to that of spherical particles. The length of the fibers only appears to have a minor influence on the entrainment threshold. No preferential alignment is observed for non-Brownian fibers on a flat plate, except for very thin films, whereas the fibers tend to align themselves along the axis of a cylindrical rod for a large enough ratio of the fiber length to the radius of the cylindrical rod. The Landau-Levich-Derjaguin is recovered for more concentrated suspension by introducing an effective capillary number accounting for the change in viscosity.} \\

 \end{@twocolumnfalse} \vspace{0.6cm}

  ]

\makeatletter
{
\blfootnote{\textit{$^{a}$~Department of Mechanical Engineering, University of California, Santa Barbara, California 93106, USA}}
\blfootnote{\textit{$^{b}$~PMMH, CNRS UMR-7636, ESPCI, PSL University, 75005 Paris, France}}
\blfootnote{\textit{$^{*}$ asauret@ucsb.edu}}
\makeatother


\section{Introduction}

Dip-coating is a widely used manufacturing process consisting of immersing a substrate in a liquid bath and withdrawing it at a prescribed velocity, which leads to the entrainment of a thin liquid layer coating its surface.\cite{scriven1988physics,weinstein2004coating} The pioneering works of Landau \& Levich\cite{landau1942dragging} and Derjaguin\cite{derjaguin1943} have demonstrated that the thickness of the film coating a flat substrate in the case of a homogeneous Newtonian liquid without evaporation depends on the physical properties of the liquid (density $\rho$, dynamic viscosity $\eta$, air-liquid surface tension $\gamma$), and on the withdrawal velocity $U$. The thickness $h$ of the film is captured through the capillary length, $\ell_c=\sqrt{\gamma/(\rho\,g)}$, where $g$ is the acceleration of gravity, and the capillary number, ${\rm Ca}=\eta\,U/\gamma$, which describes the ratio of the viscous forces and capillary effects:
\begin{equation}
h=0.944\,\ell_c\,{\rm Ca}^{2/3}. 
\label{eq:LLD}
\end{equation}
This theoretical law, known as the Landau-Levich-Dejarguin (LLD) law, has been verified experimentally and numerically.\cite{lee1974meniscus,de1998gravity,maleki2011landau,colosqui2013hydrodynamically} In addition, different parameters influencing the dip coating process have been considered: the influence of the substrate geometry, for example, rods and wires,\cite{white1966theory,quere1999fluid,ruckenstein2002scaling,zhang2022dip} the role of surface roughness,\cite{benkreira2004effect,seiwert2011coating} soft substrate,\cite{bertin2022enhanced} surfactants,\cite{park1991effects,krechetnikov2005experimental} and also more complex rheologies of the fluids used.\cite{ro1995viscoelastic,maillard2016dip,smit2019stress}


These experiments have mainly considered pure homogeneous fluids. Nevertheless, some studies have considered the dip-coating process with particle suspensions to functionalize substrates, for example, to give them optical properties, such as transparency at specific wavelengths,\cite{koh2006situ} or hydrophobic properties.\cite{matin2019uv} In the case of dip coating with colloidal suspensions, the approach is to disperse the particles to be deposited in the liquid and then coat the surface \textit{via} dip-coating by letting the solvent evaporate. The process is complex as it involves three distinct competitive mechanisms: the interactions between the liquid and the solid through a dynamic contact line, the evaporation of the solvent, and the deposition of particles on the surface.\cite{berteloot2013dip} One of the key parameters is the withdrawal velocity of the substrate compared to the evaporation rate of the liquid. 

Recently, different studies have considered the interactions between particles and capillary flows in non-volatile solvents, for instance, during the formation of drops,\cite{bonnoit2012accelerated,chateau2018pinch,thievenaz2021droplet,thievenaz2021pinch,thievenaz2022onset} of jets,\cite{furbank2004experimental,chateau2019breakup} or during the atomization of suspensions.\cite{raux2020spreading} The complexity in these systems arises when the length scale of the capillary object becomes comparable to the length scale of the particles, for instance, their diameter in the case of spherical particles. Of particular interest in manufacturing processes is the dip coating of suspensions of particles and the resulting coating films obtained.\cite{kao2012spinodal,gans2019dip,sauret2019capillary,palma2019dip,jeong2022dip} These studies have shown that the ratio of the film thickness $h$ to the particle diameter $d$ controls the transition between different coating regimes where particles are entrained on the substrate or remain in the liquid bath. More specifically, the particles remain trapped by the meniscus for low film thicknesses, and the substrate is then devoid of particles. For a dilute suspension of non-Brownian spherical particles, the threshold for a particle of diameter $d$ to be entrained in the coating film has been shown to be $h \gtrsim d/6$ both for flat \cite{sauret2019capillary} and cylindrical substrates.\cite{dincau2020entrainment} For non-dilute suspension and in the case of thick films ($h \gtrsim d$), the coating thickness can be predicted using the bulk viscosity of the particulate suspension $\eta(\phi)$, where $\phi$ is the volume fraction of particles.\cite{gans2019dip,palma2019dip} The interplay between the particle size and the film thickness on the entrainment regime has allowed the development of a capillary sorting method for polydisperse suspensions.\cite{dincau2019capillary,jeong2022dip}

The interplay between interfacial dynamics and particles can become even more challenging to describe with anisotropic particles, such as suspensions of fibers,\cite{pittman1990extensional,chateau2021extensional} An additional complexity with suspensions of fibers is that they are characterized by two dimensions: the diameter $d$ and the length $L$, leading to the definition of the aspect ratio $a=L/d$.  In the case of dip coating, it remains unclear how the entrainment threshold and the different coating regimes reported for suspensions of spherical particles translate to situations where the particles to be deposited are of more complex shapes. Indeed, the anisotropy of fibers could potentially modify their entrainment but also their alignment on a coated substrate, which remains a significant challenge in some manufacturing processes, for instance with nanowires \cite{huang2001directed,hu2017spray} or with carbon nanotubes (CNT).\cite{goh2019directed,jambhulkar2020scalable} The difference in fiber alignment during the dip-coating of wires by carbon nanotubes depending on the withdrawal direction has been qualitatively observed by Sponitz \textit{et al.}\cite{spotnitz2004dip} However, the threshold for the entrainment of fibers and their alignment, as well as the role of the substrate geometry, remain elusive.

The rheology of fiber suspensions is also more complex to describe than the rheology of suspensions of spherical particles.\cite{guazzelli2018rheology,butler2018microstructural} Indeed, in addition to the volume fraction $\phi$, the viscosity of fiber suspensions also depends on the aspect ratio $a$. Different regimes have been identified depending on $a=L/d$ and $\phi$. The separation between the different regimes (dilute, semi-dilute, and concentrated) is governed by the number density $n$ defined as the number of particles per unit volume, $n=N/V=\phi/V_{\rm p}$, where $V_{\rm p}=\pi\,d^2\,L/4$ is the volume of a fiber. This number can be made dimensionless by multiplying it by the particle dimensions, for instance, $L^3$. Dilute suspensions of fibers, for which the interactions between the fibers can be neglected, correspond to the condition $n\,L^3 \ll 1$.\cite{batchelor1971} When increasing the volume fraction of fibers, the behavior is more complex to describe as the interactions between particles become more frequent. For larger fiber volume fraction, the suspension is in a semi-dilute regime when $1 \leq n\,L^3 \ll L/d$, in which interactions between fibers start to influence the rheology. As the volume fraction of fibers increases further, rotation of the suspended fibers will be greatly limited when the number density becomes larger than $1/L^2 d$.\cite{butler2018microstructural} For denser suspension of fibers, different empirical rheological models have been proposed, and the viscosity appears to depend on the distance to the jamming point where the viscosity diverges at a maximum packing fraction $\phi_{\textrm{c}}$, which depends on the aspect ratio $a$.\cite{tapia2017rheology,bounoua2019shear} In addition to the volume fraction $\phi$, the alignment of fibers also contributes to the rheology of fiber suspensions. Under shear flow, the suspended fibers may rotate. As the strain rate of the flow determines the orientation alignment of the suspended fibers, the rheology of fiber suspension depends on the flow characteristics.\cite{folgar1984orientation, chun2023universal}


\begin{figure*}
\centering
  \includegraphics[width=0.95\textwidth]{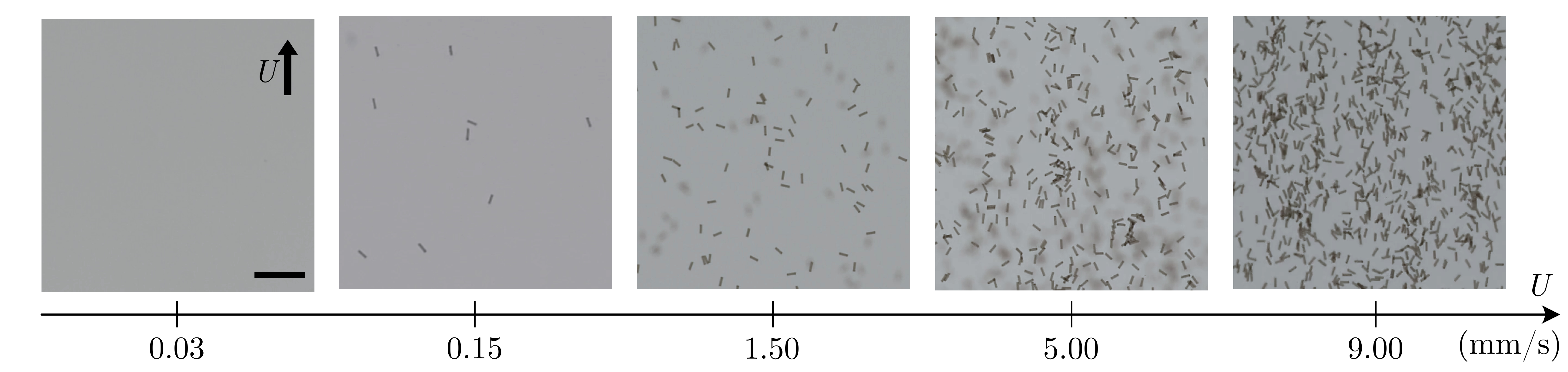}
  \caption{Examples of coating films resulting from the dip coating of a flat plate withdrawn from a suspension of non-Brownian fibers when increasing the withdrawal velocity $U$ for fibers of diameter $d=50\,\mu \textrm{m}$, and length $L=200\,\mu \textrm{m}$. The scale bar is $1\, \textrm{mm}$. The withdrawal velocity $U$, and thus the thickness $h$ of the coating film, increases from left to right. The arrow in the first picture of the figure indicates the direction along which the substrate has been withdrawn.}
  \label{fgr:deposition_illustration}
\end{figure*}

As discussed above, the complexity of describing fiber suspensions may introduce additional effects during the dip coating process. We show in figure \ref{fgr:deposition_illustration} an example of coating films observed after withdrawing a glass plate from a suspension of non-Brownian fibers. Similarly to spherical particles, different regimes are visible. For low withdrawal velocity, fibers are not entrained in the coating film. When increasing the withdrawal velocity, fibers start to be entrained in the coating film, and their number density increases with the withdrawal velocity, and hence with the film thickness. We also observe that the fibers do not exhibit preferential orientation on a planar substrate.

This study aims to characterize the dip coating of suspensions of non-Brownian fibers on different substrates (flat plates and cylindrical rods). We first present in section \ref{sec:exp} the experimental setup, the properties of the fiber suspensions used in this study, and the methods of characterization of the coated substrate. We then consider dilute suspensions of fibers, so that at first order, every fiber approaching the meniscus can be considered as isolated. We characterize the entrainment threshold on planar substrates and on cylindrical rods in section \ref{sec:entrainment} and compare our results to models developed for spherical particles.\cite{sauret2019capillary,dincau2020entrainment} Section \ref{sec:orientation} is devoted to the characterization of the orientation of the fibers entrained under different conditions. Finally, we consider in section \ref{sec:concentrated} the case of non-dilute suspensions of fibers. The results presented in this paper further extend our understanding of the dip coating of non-Brownian fiber suspensions and provide general guidelines in manufacturing processes.


\section{Experimental Methods} \label{sec:Experiments}

\section{Experimental methods} \label{sec:exp}

\begin{figure}
\centering
  \includegraphics[width=0.49\textwidth]{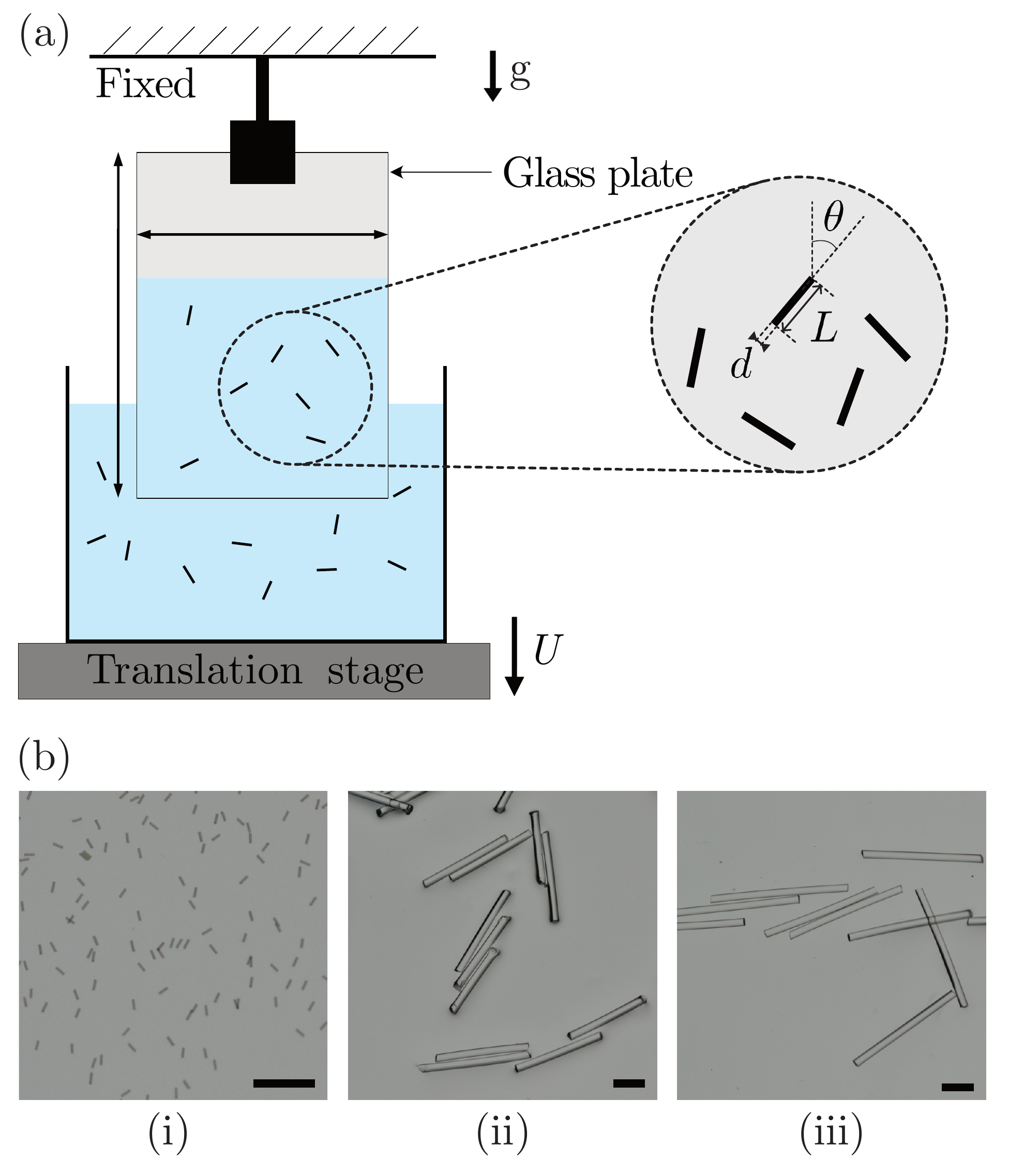}
  \caption{(a) Schematic of the experimental setup. The zoom view shows the notations used to characterize the orientation of the fibers: $\theta$ is the absolute value of the angle between the long axis of the fiber and the withdrawal direction. (b) Pictures of some fibers used in the experiments: (i) $d=50 \,\mu{\rm m}$ and $L=0.2\,{\rm mm}$, (ii) $d=280 \,\mu{\rm m}$ and $L=3\,{\rm mm}$, (iii) $d=280 \,\mu{\rm m}$ and $L=5\,{\rm mm}$. Scale bars are $1\,{\rm mm}.$}
  \label{fgr:experimental_setup}
\end{figure}

The experiments consist in withdrawing either a glass plate ($W=51\, \textrm{mm}$ wide and $e=3.2 \,\textrm{mm}$ thick, from McMaster-Carr) from a rectangular container (width $78\, \textrm{mm}$, thickness $36\, \textrm{mm}$, and height $75 \, \textrm{mm}$), or a glass capillary rod (radius $R=275$, and $600\,\mu {\rm m}$, from Vitrocom) from a cylindrical container (diameter $45 \,{\rm mm}$ and height $70 \, \textrm{mm}$) filled with a suspension of non-Brownian fibers. Figure \ref{fgr:experimental_setup}(a) shows a schematic of the experimental setup in the case of a flat plate.

The suspensions are prepared by dispersing nylon fibers in silicone oil at a volume fraction $\phi=V_{\rm p}/(V_{\rm p}+V_{\rm l})$, where $V_{\rm p}$ and $V_{\rm l}$ are the volume of fibers and interstitial liquid, respectively. Nylon fibers of density $\rho \simeq 1130\,{\rm kg.m^{-3}}$ of various diameters and lengths are used (see Table \ref{table:fibers} and examples in figure \ref{fgr:experimental_setup}(b)). The fibers with the smallest diameter ($d=50\,\mu{\rm m}$, purchased from Cellusuede Products Inc.) are received cut at the desired length, whereas the fibers of larger diameter ($d=130,\,200,\,280\,\mu{\rm m}$) are hand cut at the desired lengths from nylon wire (Berkley Trilene). The fibers are dispersed in silicone oil (Fluorosilicone FMS-221 from Gelest) of shear viscosity $\eta_0 = 109 \,\rm{mPa.s}$, density $\rho = 1160 \,\rm{kg.m^{-3}}$ and a surface tension of $\gamma=21\,{\rm mN.m^{-1}}$ at $20^{\rm o}$C. The density of the silicone oil is chosen close to the density of the fibers to minimize buoyancy effects, which can be neglected over the duration of an experiment, typically of the order of tens of seconds. In addition, silicone oil is a fluid commonly used for dip coating experiments since it perfectly wets the glass substrate and the particles and avoids any potential surfactant effects that could lead to thicker coating films even at low concentrations.\cite{krechetnikov2005experimental,krechetnikov2006surfactant,rio2017withdrawing} Between each experiment, the glass substrate is cleaned with isopropyl alcohol (Sigma-Aldrich), rinsed with DI water, and dried with compressed air.

\begin{table}[]\centering
\begin{tabular}{|lll|}
\hline
$d\,(\mu\mathrm{m})$ & $L\,(\mathrm{mm})$              & $a=L/d$         \\ \hline
$50$     & $0.1, 0.2, 0.4$    & $2, 4, 8$ \\
$130$    & $3$                   & $23$          \\
$200$    & $3$                   & $25$ \\
$280$    & $3, 5, 10$            & $11, 18, 36$  \\ \hline
\end{tabular}
    \caption{Properties of the fibers used in this study: diameter $d$, length $L$, and aspect ratio $a=L/d$.}
    \label{table:fibers}
\end{table}

The bath containing the fiber suspension is placed on a motorized stage that is translated vertically using a stepper motor (Thorlabs NRT150) at a velocity $0.1\,{\rm mm.s^{-1}}<U<10\,{\rm mm.s^{-1}}$, whereas the substrate remains fixed in the laboratory frame. The measurements of the coating thickness are done using gravimetric methods.\cite{krechetnikov2005experimental,gans2019dip} Recordings of the coating film during and at the end of the process are taken using a DSLR camera (Nikon D5600) equipped with a macro lens (Nikkor 200 mm). Examples of pictures at the end of the withdrawal are shown in figure \ref{fgr:deposition_illustration}. Using a custom-made ImageJ routine,\cite{abramoff2004image} we extract the number of deposited fibers from these pictures, as well as their position and orientation $\theta$.


\section{Entrainment threshold of fibers} \label{sec:entrainment}

Particle entrainment in the liquid film during the withdrawal of a substrate from a suspension bath is due to the interaction of capillary, viscous, and frictional forces on a particle at the meniscus. The viscous and frictional forces drag the particle into the coating film, while the capillary force pushes it back into the bulk of liquid.\cite{jeong2020deposition} When increasing the withdrawal velocity $U$, the thickness $h$ of the film and at the stagnation point $h^*=3\,h$ increase. When the film thickness becomes larger than a fraction of the diameter of a spherical particle, isolated particles start to be entrained in the coating film.\cite{sauret2019capillary}

In this section, we characterize the deposition threshold of fibers. We investigate how the diameter $d$ and the length $L$ of a fiber, affect the entrainment threshold, first in the case of a planar surface, and then for a cylindrical substrate. All the experiments performed here are done at very small volume fractions ($\phi <0.5\%$), and the suspensions are in the dilute regime ($nL^3 \ll 1$.)

\subsection{Phenomenology}

We report in figures \ref{fgr:Figure_4}(a)-(c) examples of the evolution of the coating film when increasing the withdrawal velocity $U$, and hence the film thickness $h$. For a given size of fibers, the coating film is devoid of particles at low withdrawal velocity [figure \ref{fgr:Figure_4}(a)]. Then at a certain larger withdrawal velocity, clusters of fibers start to be entrained in the coating film, but no isolated fibers are observed [figure \ref{fgr:Figure_4}(b)]. Finally, beyond a threshold withdrawal velocity, $U^*$, individual fibers are entrained in the film [figure \ref{fgr:Figure_4}(c)]. Increasing further the withdrawal velocity increases the number density of fibers deposited on the plate. For dilute suspensions of fibers, \textit{i.e.}, very small volume fraction $\phi$, the threshold value of the entrainment of isolated fibers does not depend on $\phi$. The different regimes observed here are qualitatively similar to the one reported for spherical particles on a planar substrate.\cite{sauret2019capillary}

We first consider how a single fiber is entrained in the coating film, as shown by the example reported in figure \ref{fgr:Figure_4}(c). In the case of a coating film thickness smaller than a fraction of the fiber diameter, we observe that when an isolated fiber arrives at the air-liquid meniscus, the fiber rotates, and its principal axis aligns with the meniscus due to capillary force.\cite{lewandowski2008rotation,cavallaro2011curvature} After that, the fiber starts rotating when entering the coating film, one of its extremities staying pinned to the meniscus [figure \ref{fgr:Figure_4}(c), right]. This effect is due to the shape of the fibers and some potential small defects at the edges of the fibers during their manufacturing. Indeed, the exact local diameter at the two edges of a fiber can be slightly different leading to the larger edge remaining pinned at the meniscus and the fiber rotating until the drag and friction forces are sufficient to entrain the fiber. 

In the following, we consider the influence of the fiber diameter and length on the entrainment threshold, first for a planar substrate, and then for a cylindrical one. 

\begin{figure}[h!]
\centering
  \includegraphics[width=0.495\textwidth]{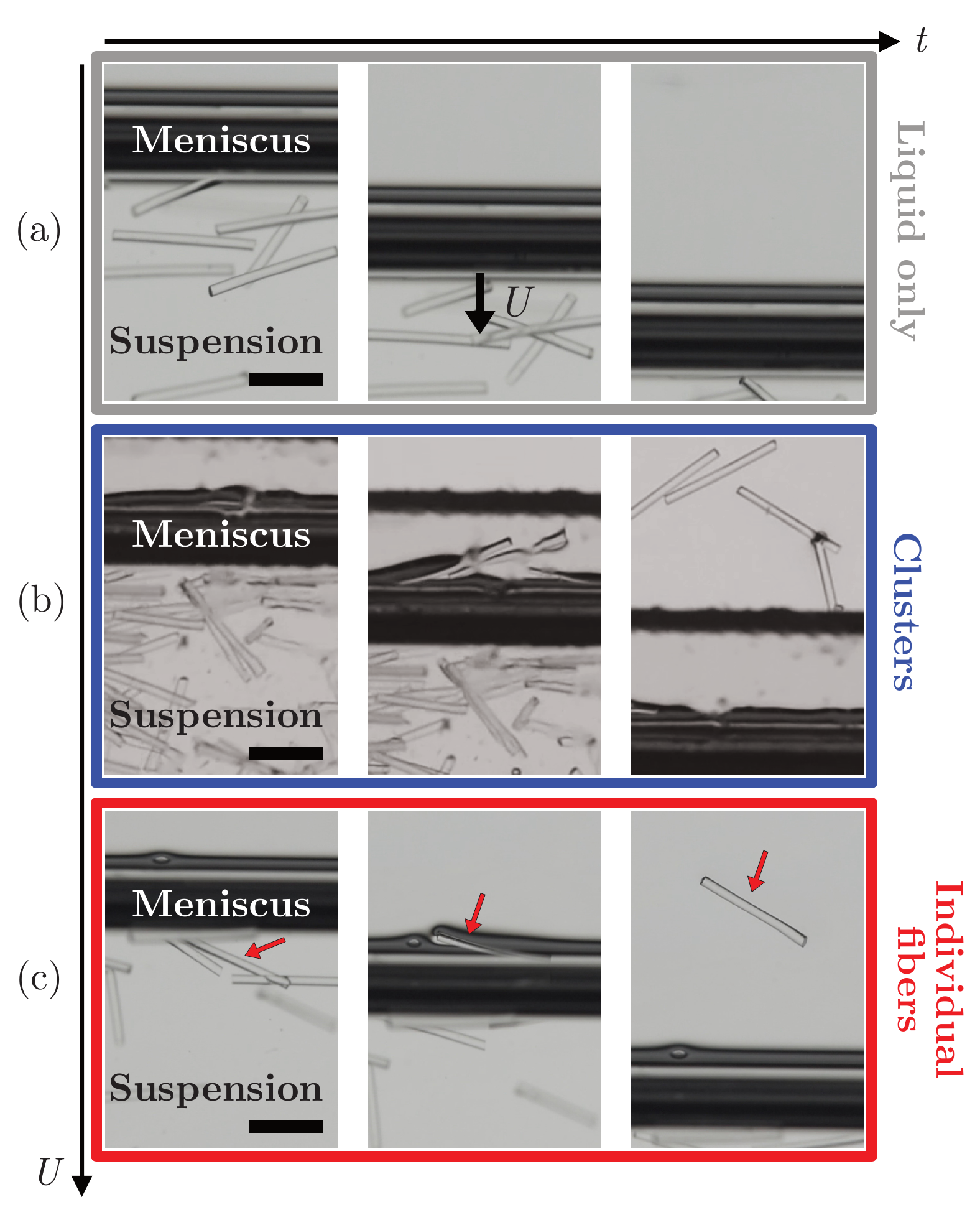}
  \caption{Pictures of the deposition process for the three coating regimes obtained at different withdrawal velocities $U$ for fibers of diameter $d=280\,\mu\mathrm{m}$ and length $L=3\,{\rm mm}$. (a) At very low velocity, here $U=0.1\mathrm{mm/s}$, only the liquid is entrained on the substrate and the fibers remain trapped in the suspension bath. (b) For intermediate velocities, here $U=1\mathrm{mm/s}$, clusters of fibers start to be entrained on the substrate. (c) At larger withdrawal velocity, here $U=9\mathrm{mm/s}$, individual fibers are able to be entrained in the coating film. Scale bars are $2\,\mathrm{mm}$.}
  \label{fgr:Figure_4}
\end{figure}

\subsection{Planar substrate}

We first consider the case of a planar substrate (glass plate) withdrawn from the suspension bath. The inset of figure \ref{fgr:Figure_5} shows the threshold withdrawal velocity $U^*$ at which individual fibers start to be entrained in the coating film when the glass plate is withdrawn from the suspension. We observe that the diameter $d$ of the fibers has a strong influence on the entrainment threshold. The larger the diameter, the larger the threshold withdrawal velocity is. On the other hand, for a given diameter, $d=280\,\mu{\rm m}$ here, changing the length of the fiber from $L=3\,{\rm mm}$ to $L=10\,{\rm mm}$ suggests that the threshold entrainment velocity $U^*$ increases slightly with the length of the fiber. Nevertheless, the length of the fibers has a weaker influence on the threshold velocity than its diameter.

Past studies have shown that non-Brownian spherical particles can be individually entrained in the coating film if their diameter $d$ is a fraction of the thickness of the liquid film at the stagnation point $h^*$:\cite{sauret2019capillary,dincau2020entrainment} $h^*\leq \alpha\,d/2$, where $\alpha$ is close to $1$,\cite{jeong2020deposition} and shown experimentally to be $\alpha=1.15\pm 0.1$. 

\begin{figure}
\centering
  \includegraphics[width=0.48\textwidth]{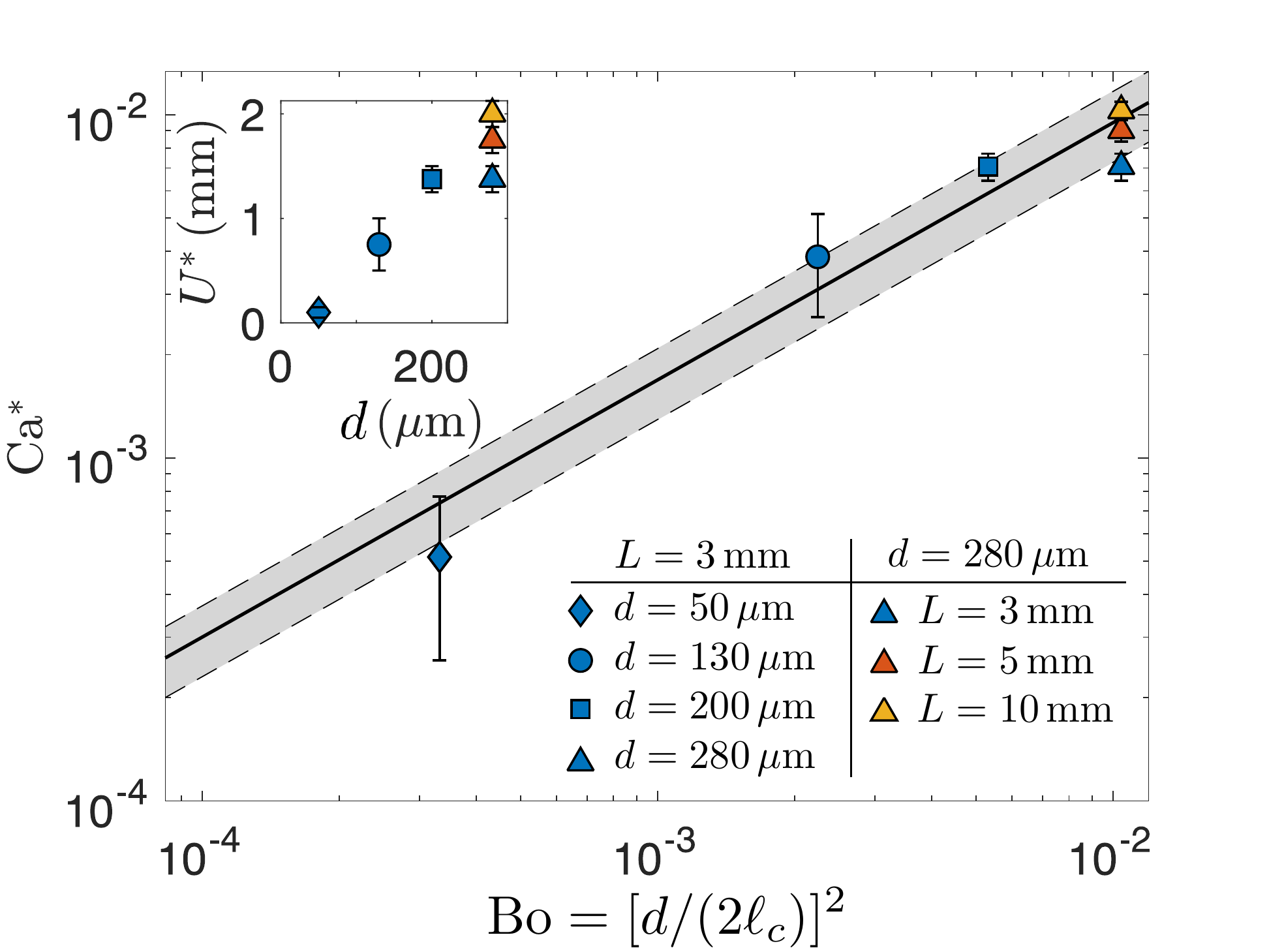}
  \caption{Threshold capillary number $\mathrm{Ca}^*$ for the entrainment of fibers on a flat substrate as a function of the Bond number $\mathrm{Bo}=[d/(2\,\ell_{\rm c})]^2$. The solid line corresponds to Eq. (\ref{eq:scaling_law}) with $\beta = 0.30 \pm 0.07$, and the grey shaded region is the uncertainty. Inset: Threshold withdrawal velocity $U^*$ for a planar substrate as a function of the fiber diameter $d$.
  }
  \label{fgr:Figure_5}
\end{figure}

Using the approach developed for spherical particles, and considering that the diameter of the fibers is the relevant length scale that captures the entrainment threshold, we introduce the Bond number as $\mathrm{Bo}=[d/(2\,\ell_{\rm c})]^2$, where $d$ is the diameter of the fiber and $\ell_{\rm c}$ the capillary length. The threshold capillary number at which isolated fibers start to be entrained in the liquid film is $\mathrm{Ca}^*=\eta_0\,U^*/\gamma$. We report the experimental results in the main panel of figure \ref{fgr:Figure_5} for different fiber diameters and fiber lengths. We observe that the experimental results collapse well on a line corresponding to the scaling law $\mathrm{Ca}^* \propto \mathrm{Bo}^{3 / 4}$.

More quantitatively, a spherical particle of diameter $d$ is entrained in the coating film if ${\rm Ca}>{\rm Ca^*}$ where:\cite{sauret2019capillary}
\begin{equation}
    {\rm Ca^*}=\beta\,\left(\frac{d}{2\,\ell_c}\right)^{3/2} =\beta\,\mathrm{Bo}^{3/4} \quad {\rm with} \quad \beta= 0.24.
    \label{eq:scaling_law}
\end{equation}

We report this prediction for the entrainment of fibers in figure \ref{fgr:Figure_5}, using the diameter of the fiber $d$ as the relevant length scale. The experimental data collapse well on this prediction with a prefactor $\beta= 0.30 \pm 0.07$, compatible with the prefactor $\beta= 0.24$ reported for spherical particles. In addition, for a given diameter, $d=280\,\mu{\rm m}$ here, different lengths of fibers also lead to entrainment thresholds of the same order of magnitude. The variations observed indicate that the longer the fiber is, the larger the threshold capillary number for entrainment is. However, the influence of the length is weaker and is of second order compared to the role of the fiber diameter. 

In summary, the entrainment threshold of an isolated fiber on a flat substrate can be predicted at first order using the diameter of the fibers, and the threshold capillary number $\mathrm{Ca}^*$ is mainly controlled by the Bond number $\mathrm{Bo}=[d/(2\,\ell_{\rm c})]^2$. The length of the fiber $L$ has a second-order role, as longer fibers need a slightly larger withdrawal velocity to be entrained on a planar substrate.

\subsection{Cylindrical substrate}
During the dip coating of cylindrical substrates, \textit{i.e.}, rods, the radius of the rod $R$ adds an additional curvature.\cite{quere1999fluid} The thickness of the coating film depends on the ratio of the radius of the substrate $R$ and the capillary length $\ell_c$, captured through the Goucher number, ${\mathrm{Go}}=R/\ell_c$.\cite{rio2017withdrawing} Different expressions have been proposed for the thickness $h$ of the deposited liquid film on a cylindrical rod at non-zero Goucher number by matching the flowing film curvature to the static curvature,\cite{white1966theory,quere1999fluid} In the thin-film limit, Dincau \textit{et al.} \cite{dincau2020entrainment} simplified the approach of White \& Tallmadge\cite{white1966theory} and provided an explicit expression of the film thickness $h$ as a function of the capillary number ${\rm Ca}$ and the Goucher number ${\rm Go}$:
\begin{equation} \label{eq:LLD_cylinder_simplified}
\frac{h}{R}=\frac{1.34 \mathrm{Ca}^{2 / 3}}{1+2.53 \mathrm{Go}^{1.85} /\left[1+1.79 \mathrm{Go}^{0.85}\right]},
\end{equation}
which captures quantitatively the evolution of $h$ with $\mathrm{Ca}$ for cylindrical rods of finite radius. For dilute suspensions of spherical particles, and similarly to a flat substrate, the thickness at the stagnation point $h^*$ has to be larger than a fraction of the particle diameter $d$ to entrain the particle in the coating film: $h^*\geq \alpha d/2$, where $1 \leq \alpha \leq 2$. For spherical particles, Dincau \textit{et al.}\cite{dincau2020entrainment} have shown that the condition $h^*\geq \alpha d/2$ together with $h^*=3\,h$ leads to an equation for $\mathrm{Ca}^*$:
\begin{equation}
\frac{\alpha d}{6 R}=\frac{1.34 \mathrm{Ca}^{* 2 / 3}}{1+2.53 \mathrm{Go}^{1.85} /\left[1+1.79 \mathrm{Go}^{0.85}\right]},
\end{equation}
and further rearranging for the threshold capillary number leads to:
\begin{equation} \label{eq:fiber_thresholdCa}
\mathrm{Ca}^* \simeq 0.645\left[\frac{\alpha d}{6 R}\left(1+\frac{2.53 \mathrm{Go}^{1.85}}{1+1.79 \mathrm{Go}^{0.85}}\right)\right]^{3 / 2},
\end{equation}
with $\alpha \simeq 1.1 \pm 0.1$ obtained experimentally for spherical particles. 

Here, we consider cylindrical rods of radius $R=275\,\mu{\rm m}$ and $R=600\,\mu{\rm m}$ in a dilute suspension of fibers of diameter $d=50\,\mu{\rm m}$ and two lengths, $L=400\,\mu{\rm m}$ and $L=1000\,\mu{\rm m}$. The experimental threshold velocities $U^*$ at which the fibers are entrained are reported in the inset of figure \ref{fgr:Figure_6}. Similarly to the case of spherical particles,\cite{dincau2020entrainment} we observe that for a given fiber radius and length, the smaller the radius of the cylindrical rod, the larger the threshold withdrawal velocity $U^*$. Here again, the experimental results suggest that the relevant length scale to describe the entrainment of isolated fibers is, at first order, their diameter. Nevertheless, for the smallest radius of rods, the length of the fibers seems to have a larger influence than what we observed for a flat plate.

We report in the main panel of figure \ref{fgr:Figure_6} the experimental results rescaled using Eq. (\ref{eq:fiber_thresholdCa}). The entrainment threshold of fibers follows quantitatively the theoretical prediction given by Eq. (\ref{eq:fiber_thresholdCa}), especially when the aspect ratio between the length of the fibers $L$ and the radius of curvature of the substrate $R$ is small. For larger values of the aspect ratio $L/R$, \textit{i.e.}, when the fibers are a few times longer than the radius of curvature of the rod, our results suggest that a larger capillary number, hence a larger film thickness, is required to entrain fibers, as illustrated by the difference between $L=400\,\mu{\rm m}$ and $L=1000\,\mu{\rm m}$ for $R=275\,\mu{\rm m}$ in Figure \ref{fgr:Figure_6}. 

In summary, Eq. (\ref{eq:fiber_thresholdCa}), initially established for spherical particles,\cite{dincau2020entrainment} can be used to estimate the entrainment threshold for short enough fibers compared to the diameter of the substrate, $L/R \lesssim 2$ here. In this case, similarly to flat substrates, the length scale to consider is the diameter of the fibers $d$. Nevertheless, when the length of the fibers is of an order of magnitude larger than the diameter of the rod, $L/R \gtrsim 2$, fibers are entrained with their main axis perpendicular to the meniscus, whereas for a plate or for cylindrical rods of large radius, such that $L/R \lesssim 1$, we observed that the fibers could be entrained with their main axis parallel to the meniscus. The elasto-viscous number of the fiber in this situation, defined as ${\rm Sp}={16\,\eta\,U\,L^4}/(d\,^4\,E)$, where $U$ is the flow velocity, $E$ is the Young's modulus of the fiber, and $\dot{\gamma} \sim U/L$ is the shear rate, in the range $10^{-15} \lesssim {\rm Sp} \lesssim 10^{-4}$.\cite{quennouz2015transport,du2019dynamics} Since the elasto-viscous number remains very small, the fiber can be considered rigid. Thus no wrapping is possible around the cylindrical rod, and the direction of entrainment has to be overall along the withdrawal direction. As a result, the threshold capillary number increases with $L/R$.

\begin{figure}[h!]
\centering
  \includegraphics[width=0.48\textwidth]{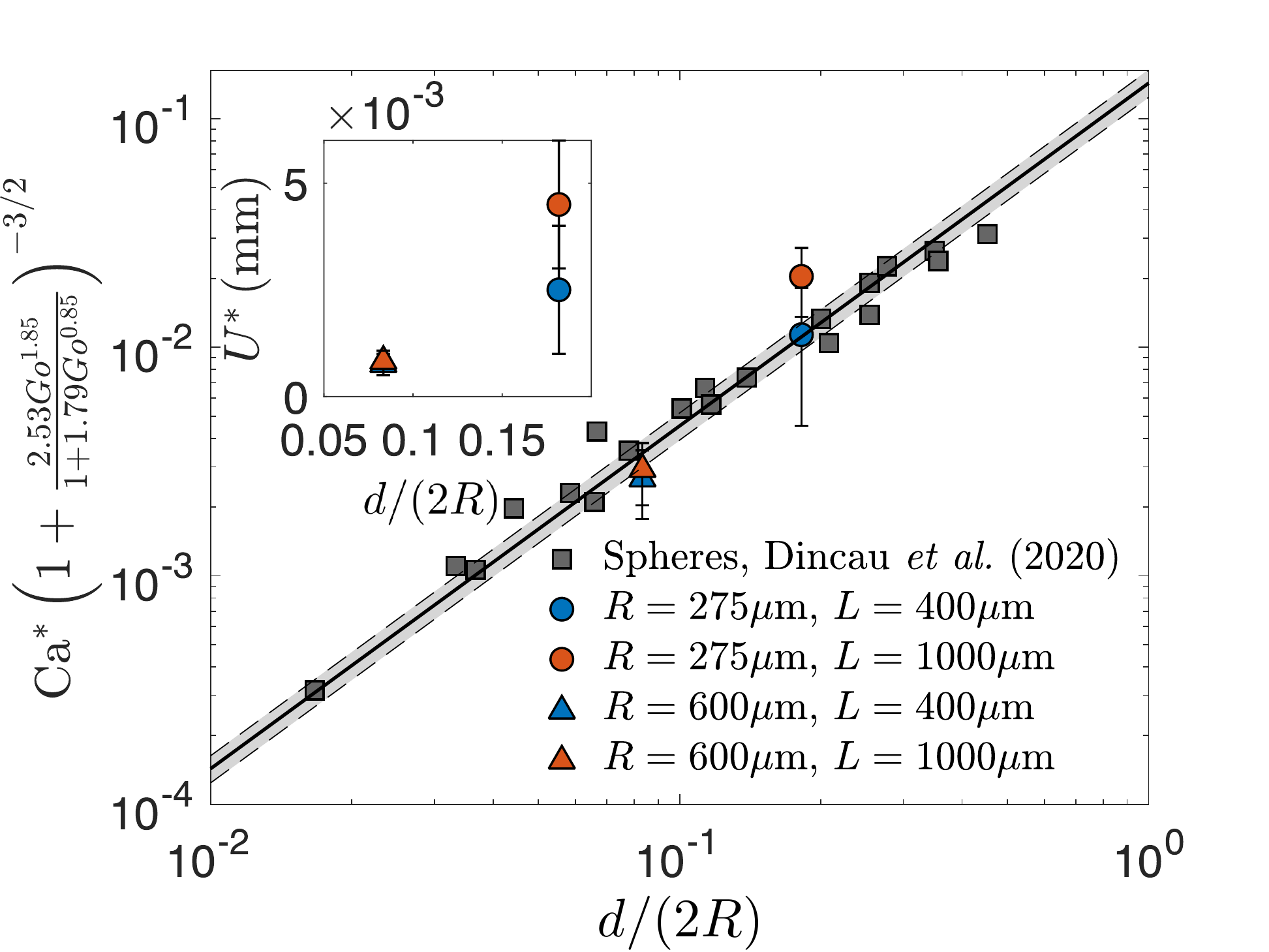}
  \caption{Rescaled threshold capillary number of entrainment of isolated fibers on cylindrical rods as a function of the ratio $d/(2\,R)$. The diameter of the fibers in the suspension is $d=50\,\mu\mathrm{m}$. The solid line corresponds to Eq. (\ref{eq:fiber_thresholdCa}) where $\alpha \simeq 1.1 \pm 0.1$. The grey-shaded region corresponds to the uncertainty on $\alpha$. Inset: Threshold withdrawal velocity $U^*$ for fiber entrainment on cylindrical rods as a function of $d/(2\,R)$.}
  \label{fgr:Figure_6}
\end{figure}

\begin{figure}
\centering
  \includegraphics[width=0.48\textwidth]{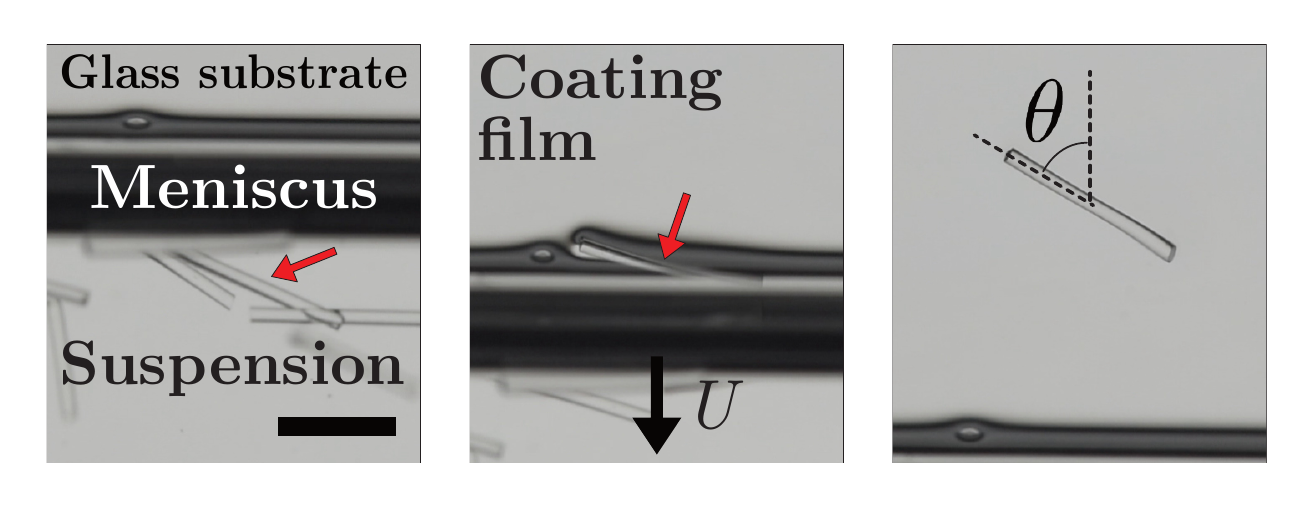}
  \caption{Illustration of the entrainment of a fiber on a glass plate showing the value of the angular orientation of the deposited fiber, $\theta$. Add: The red arrow shows the fiber depositing on the surface of the substrate. Scale bar is $2\,\mathrm{mm}$.  \label{fgr:angle}}
\end{figure}


\section{Orientation of the fibers entrained in the coating film} \label{sec:orientation}

The orientation of the deposited fibers has significant importance in practical applications.\cite{spotnitz2004dip} During the dip coating process, the fibers deposit on the substrate at a certain angle due to their anisotropy. In this section, we characterize their alignment on planar and cylindrical substrates. More specifically, we report the absolute value of the angular orientation of the fibers $\theta$, as defined in figure \ref{fgr:angle}. The value of the angle is $\theta=0^{\rm o}$ when the fiber is vertically oriented (along the withdrawal direction), and $\theta=90^{\rm o}$ for fibers aligned perpendicularly to the withdrawal direction (aligned with the meniscus). The angle $\theta$ is measured shortly after the entrainment when the fibers exit the meniscus and lie in the coating film constant thickness $h$, to reduce the influence of possible reorientations due to the potential drainage of the liquid on the substrate.\cite{keeley1988draining,buchanan2007pattern}


Figures \ref{fgr:orientation_plate2}(a)-(c) report the probability density function (PDF) of the angle $\theta$ on a planar substrate for fibers of diameter $d=50\,\mu{\rm m}$ and different lengths $L=200,\,400$, and $1000\,\mu {\rm m}$, respectively, while withdrawing the substrate at different velocities $U=0.15,\,0.5$, and $3.0\,{\rm mm/s}$. At large withdrawal velocities, such that the thickness at the stagnation point $h^*$ is larger than the fiber diameter $d$, the fibers do not exhibit any preferential orientation, leading to a mostly uniform distribution. The length of the fibers does not modify this observation. Since $h^*>d$, the fibers can cross the meniscus, are entrained in the coating film without touching the interface, and are not subject to capillary forces. In this regime, the orientation of the fibers is not controlled by their length or the withdrawal velocity. Interestingly, when $h^*$ becomes smaller than the diameter $d$ of the fibers, we observe fewer fibers showing a large angle $\theta$. More specifically, the probability of finding deposited fibers with $\theta \gtrsim 40-60^{\rm o}$ becomes smaller. Indeed, when a fiber arrives at the meniscus, and $h^* \leq d$, the fiber initially orients itself parallel to the meniscus, independently of its initial orientation in the suspension bath, and then deform the air-liquid interface to be entrained in the coating film. Due to small defects at the edges of the fiber, we typically observed that one edge of the fiber starts to be entrained while the other edge remains pinned at the meniscus, thus decreasing the angle between the fiber and the withdrawal direction. Eventually, the fiber gets entrained when the viscous, friction, and capillary forces are sufficient to fully deform the meniscus and the fiber deposits in the coating film. As a result of this dynamics, the PDF of the angle $\theta$ is smaller for large values of $\theta$. Here again, the length of the fiber does not appear to play a significant role.

\begin{figure}[h!]
\centering
  \includegraphics[width=0.48\textwidth]{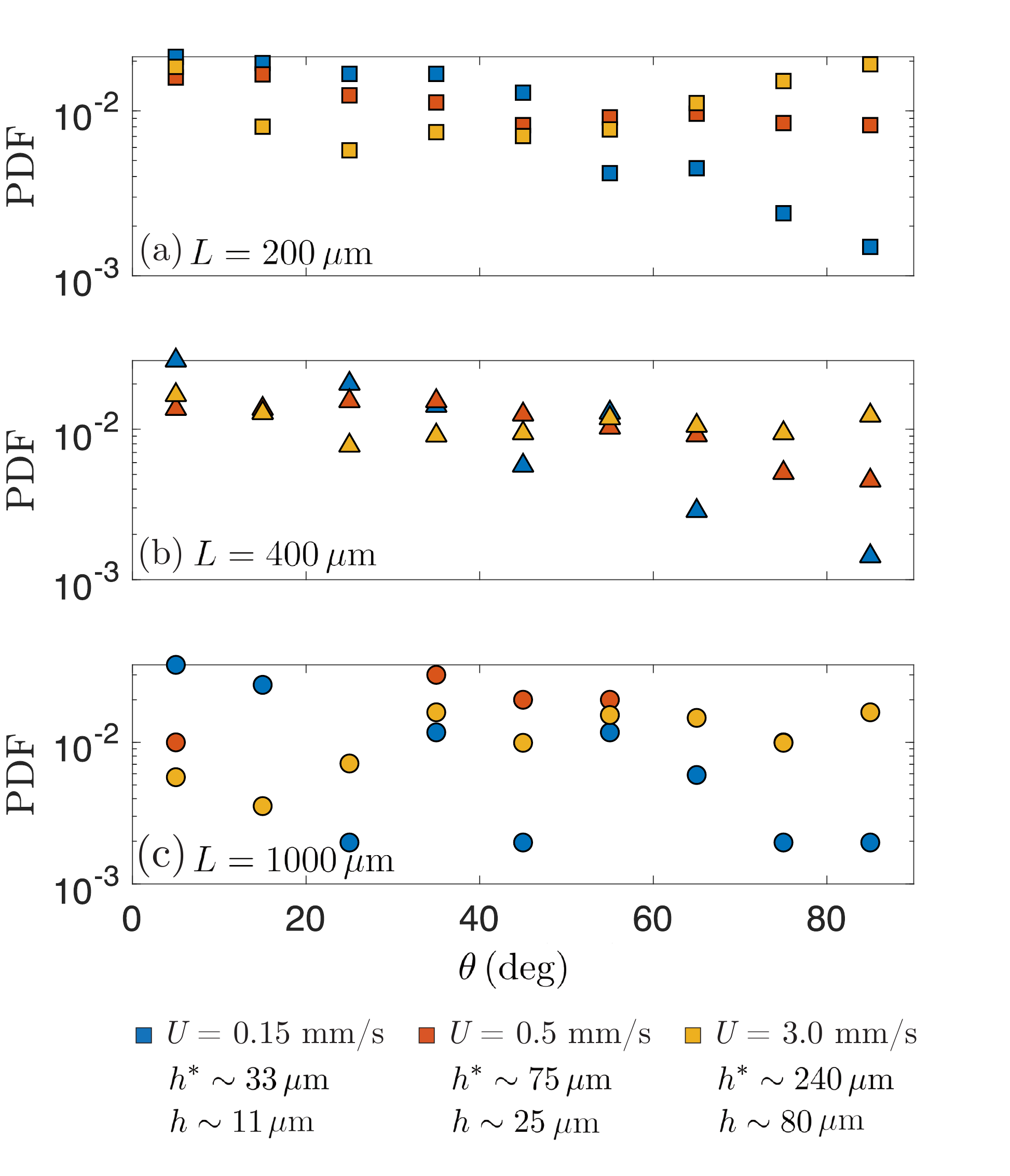}
  \caption{Probability density function (PDF) of the orientation of fibers deposited on a flat substrate for different withdrawal velocities and thus thickness $h$ of the coating film. The fibers have a diameter $d=50\,\mu\mathrm{m}$ and different lengths: (a) $L=200\,\mu\mathrm{m}$, (b) $L=400\,\mu\mathrm{m}$, and (c) $L=1000\,\mu\mathrm{m}$. \label{fgr:orientation_plate2}}
\end{figure}


We then considered the orientation of the fibers entrained on a cylindrical rod, which introduces an additional length scale, the radius of curvature of the substrate $R$. We show in figures \ref{fgr:orientation_rod}(a)-(c) the PDF of the orientation of the deposited fibers for aspect ratios of the length of the fibers to the radius of curvature of the substrate $L/R=3.6,\,1.5$ and $\,0.7$, respectively. In figures \ref{fgr:orientation_rod}(a) and \ref{fgr:orientation_rod}(b), \textit{i.e.}, $L/R > 1$, no fibers are deposited with $\theta \geq 30^{\rm o}$. Indeed, in this case, the fibers entrained in the liquid film are mostly aligned with the axis of the rod, \textit{i.e.}, perpendicular to the meniscus. When the aspect ratio is decreased, for instance, for $L/R=0.7$ in figures \ref{fgr:orientation_rod}(c), the fibers do not show such a strong preferential orientation. However, compared to the flat substrate described above, more fibers are found to exhibit a small value of $\theta$. The smaller the withdrawal velocity is, the more the fibers are aligned with the axis of the rod. Therefore, when a rigid fiber arrives close to the meniscus, the capillary force makes the fiber rotates and align with the axis of the rod, as it cannot wrap around. Thus, the fiber is entrained preferentially in the direction along the withdrawal direction. This effect is particularly pronounced when the length of the fiber is larger than the radius of curvature of the cylindrical rod, $L/R \gtrsim 1$. When the length of the fiber becomes smaller than $R$, and the smaller $L/R$ is, the more the situation becomes similar to the one of a flat substrate as the fiber sees locally less curvature. The limit $L/R \to 0$ indeed corresponds to the flat substrate case. The withdrawal velocity, hence the film thickness, appears to have a role [see figure \ref{fgr:orientation_rod}(c)], as observed for the flat substrate.

\begin{figure}
\centering
  \includegraphics[width=0.48\textwidth]{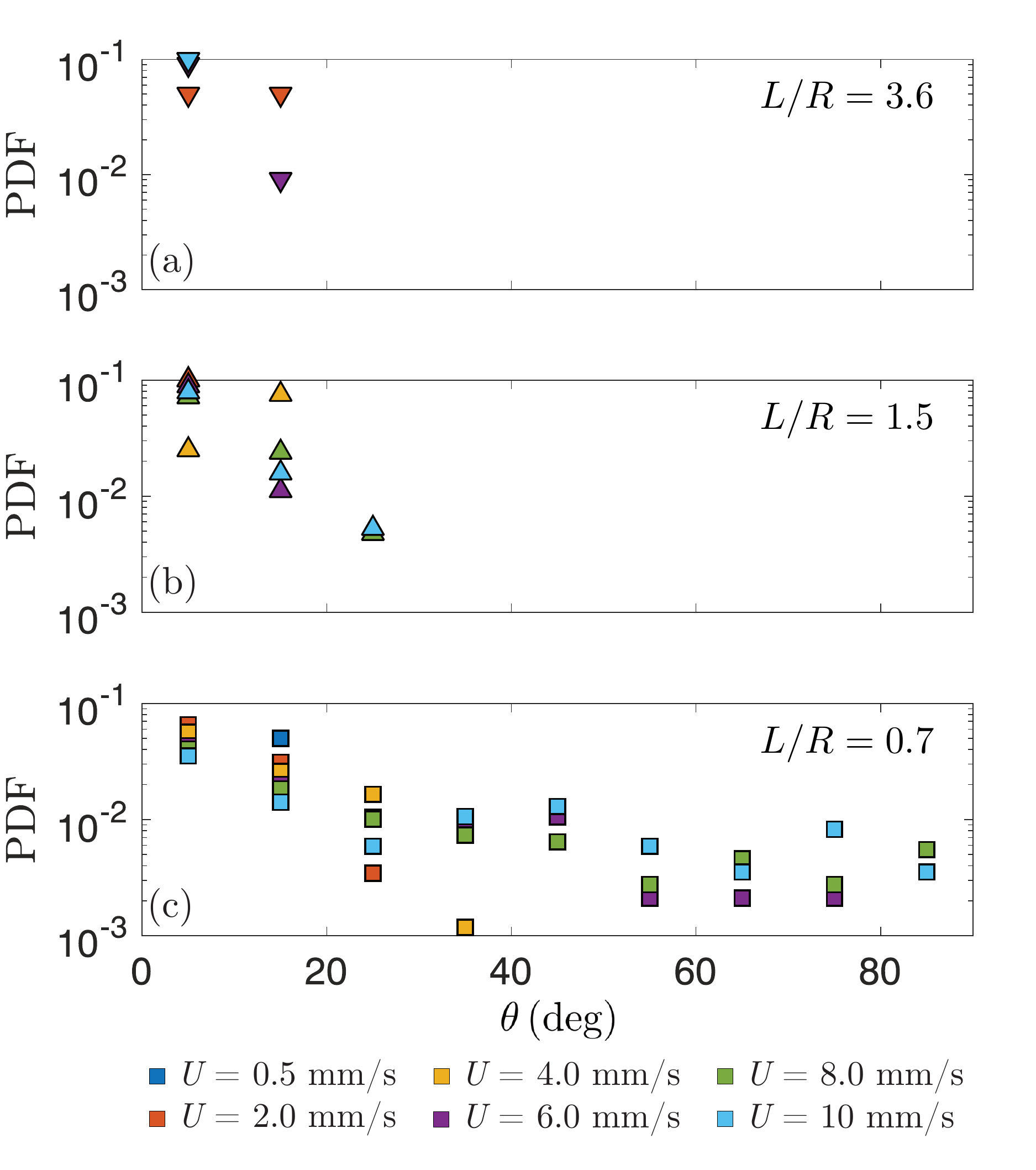}
  \caption{Probability density function (PDF) of the orientation of fibers deposited on cylindrical substrates. The radius of the cylindrical rods, the length of the deposited fibers, and the aspect ratios are (a) $R=275\,\mu \mathrm{m}$, $L=1000\,\mu \mathrm{m}$, $L/R=3.6$ (b) $R=275\,\mu \mathrm{m}$, $L=400\mu \mathrm{m}$, $L/R=1.5$ , and (c)$R=600\,\mu m$, $L=400\,\mu \mathrm{m}$, $L/R=0.7$ .}
\label{fgr:orientation_rod}
\end{figure}


We show in figure \ref{fgr:Figure_7} that the evolution of the mean absolute value of the angle of orientation $\langle \theta \rangle $ shows a monotonic behavior when increasing the aspect ratio $L/R$. For small values of $L/R$, and in particular, for a flat substrate corresponding to $L/R=0$, $\langle \theta \rangle $ is around $30^{\rm o}-45^{\rm o}$, and more importantly shows a large standard deviation due to the mainly uniform alignment between $0^{\rm o}$ and $90^{\rm o}$ of the fibers deposited on the substrate. A quite similar observation can be made for the cylindrical substrate when $L/R \lesssim 1$, although some small alignment starts to be observed. As $L/R$ increases (\textit{i.e.}, longer fibers or substrates with larger radii of curvature), $\langle \theta \rangle $ and the standard deviation decrease accordingly, showing that the fibers are aligning more and more with the axis of the cylindrical substrate. When $L/R  \gtrsim 1$, the mean absolute value of the orientation of the deposited fibers becomes $\langle \theta \rangle  \lesssim15^{\rm o}$, showing significant fiber alignment. At the highest $L/R$ considered here, $L/R=3.6$, the fibers are primarily aligned with the axis of the cylindrical rod, with $\theta \sim 0^{\rm o}$.

\begin{figure}
\centering
  \includegraphics[width=0.48\textwidth]{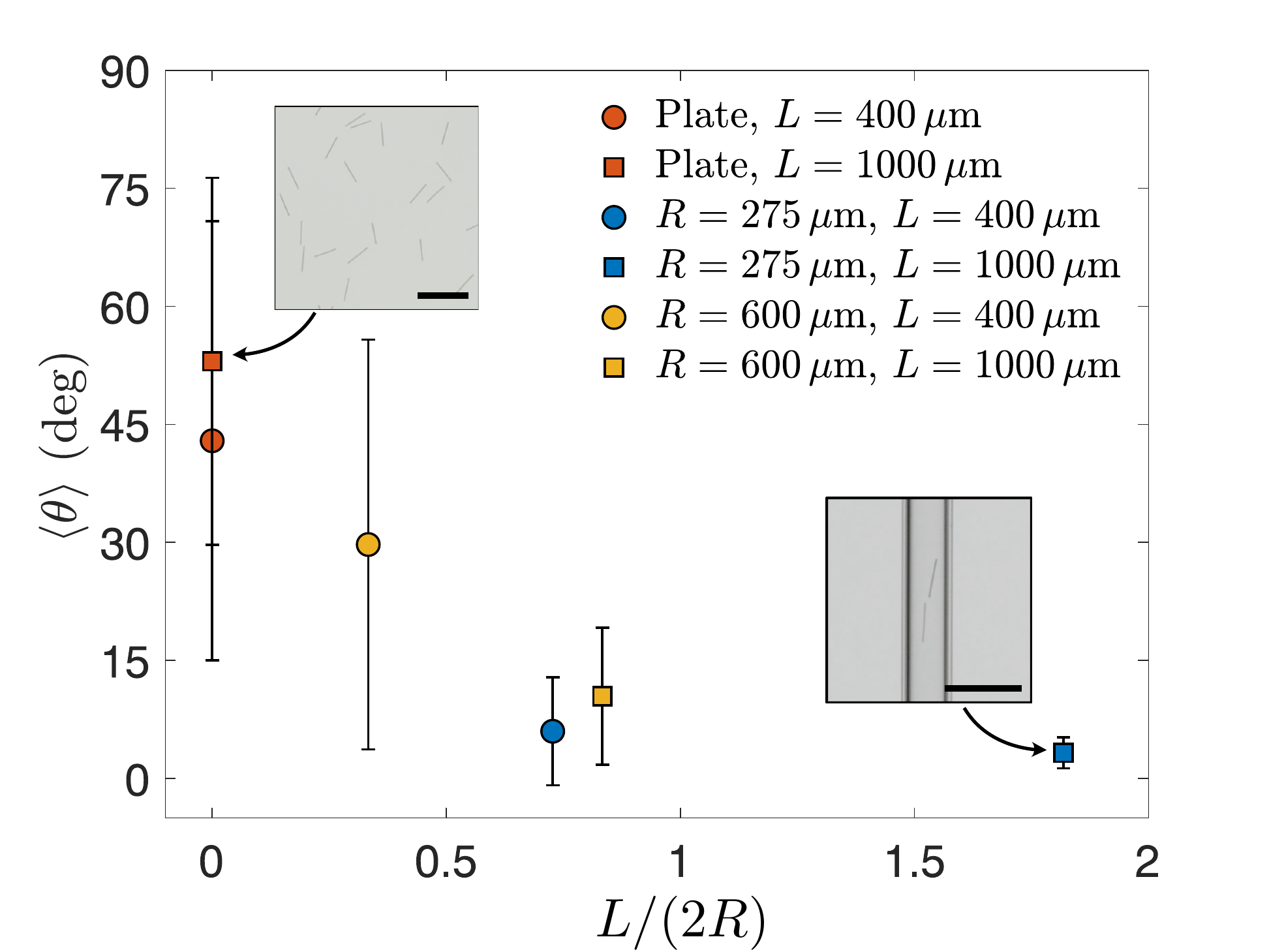}
  \caption{Mean absolute value of the angular orientation of the deposited fibers $\langle \theta \rangle$ as a function of $L/R$ for a withdrawal velocity $U=3\,\mathrm{mm}$. The fibers have a diameter $d=50\,\mu\mathrm{m}$. $L/R=0$ corresponds to flat substrates. Insets: Pictures of the deposited fibers. The dashed arrows point to the corresponding data points. Scale bars are $2\,\mathrm{mm}$.}
  \label{fgr:Figure_7}
\end{figure}

In summary, the ratio of the fiber length and the radius of curvature of the substrate $L/R$ is the key parameter that determines the alignment of fiber orientation. When a planar substrate ($L/R=0$) is withdrawn from a bath of fiber suspension, the deposited fibers show no preferential orientation, regardless of the fiber length at high velocities. Decreasing the withdrawal velocity, such that $h^* \lesssim d$ allows a slight control over the orientation of the fibers. As the ratio $L/R$ increases, the fibers start aligning with the axis of the cylindrical substrate regardless of the withdrawal velocity, and the alignment becomes significant. As a result, controlling either the length of the fibers in the suspension or the radius of the cylindrical substrates could promote the alignment of fibers during the dip-coating process.


\section{Towards denser fiber suspensions} \label{sec:concentrated}

In the previous sections, the volume fractions of fibers $\phi$ were small enough to consider that the mean spacing between fibers was large enough for the fibers not to interact significantly with each other, \textit{e.g.}, through hydrodynamic interactions or collisions.\cite{butler2018microstructural} At first order, in the limit $\phi \to 0$, the change in viscosity is negligible so that $\eta(\phi \to 0) \simeq \eta_0$. As the volume fraction of fibers increases, the viscosity of the suspension $\eta(\phi)$ is increased compared to the viscosity of the solvent. The rheology of fiber suspensions is more complex than spherical particles due to the anisotropy of fibers. Indeed, not only the volume fraction $\phi$ but also the aspect ratio of the fibers $a=L/d$ affects the rheology.\cite{butler2018microstructural} The divergence in viscosity of a suspension of fibers occurs at a value $\phi \to \phi_{\rm c}$ smaller than what is observed for spherical particles. Besides, the larger the aspect ratio $a=L/d$ is, the smaller $\phi_{\rm c}$ is. The rheology of fiber suspensions can also exhibit shear-thinning behavior when the fibers get aligned with the flow.\cite{ganani1985suspensions} Different approaches have been used to describe the viscosity of suspensions of fibers, and we here use the approach of Bounoua \textit{et al.}\cite{bounoua2019shear} who fitted the viscosity of fiber suspensions with the Maron-Pierce model :
\begin{equation}
    \label{eq:MP}
    \eta_{\mathrm{r}}(\phi) = \left( 1 - \frac{\phi}{\phi_c} \right)^{-2},
\end{equation}
where $\phi_{\rm c}$ is an empirical parameter such that the viscosity of the suspension diverges as $\phi \to \phi_{\rm c}$. We should emphasize that other possible empirical laws could have been considered; in particular, some of these rheological models present a divergence of the viscosity close to $\phi_{\rm c}$ as $(\phi_{\rm c}-\phi)^{-1}$.\cite{tapia2017rheology,khan2022rheology} Nevertheless, we use here suspensions of moderate volume fraction $\phi$ so that the difference in the prediction of the viscosity between the empirical models is not significant.

We measured the thickness of the coating film entrained on a flat substrate using a gravimetry method.\cite{krechetnikov2005experimental} We considered here suspensions with fibers of aspect ratio $L/d=4$ and volume fractions in the range $0\% \leq \phi \leq 7.5\%$. The experiments performed at larger volume fractions were challenging due to the self-filtration of concentrated fibers at the meniscus and jamming within the container when the plate was withdrawn. We report in the bottom inset of figure \ref{fgr:LLD} the evolution of $h$ when increasing the withdrawal velocity $U$. For all volume fractions used here, we observe that $h \propto U^{2/3}$ when the thick-film regime is reached, corresponding to a film thickness comparable to the particle diameter, $h \gtrsim 50\,\mu{\rm m}$.\cite{gans2019dip} This power law is in agreement with the LLD law [Eq. (\ref{eq:LLD})], and the vertical shift of the curves is a signature of the change in viscosity, as previously observed with monodisperse and bidisperse suspensions of spherical particles.\cite{jeong2022dip}

\begin{figure}
\centering
 \includegraphics[width=0.5\textwidth]{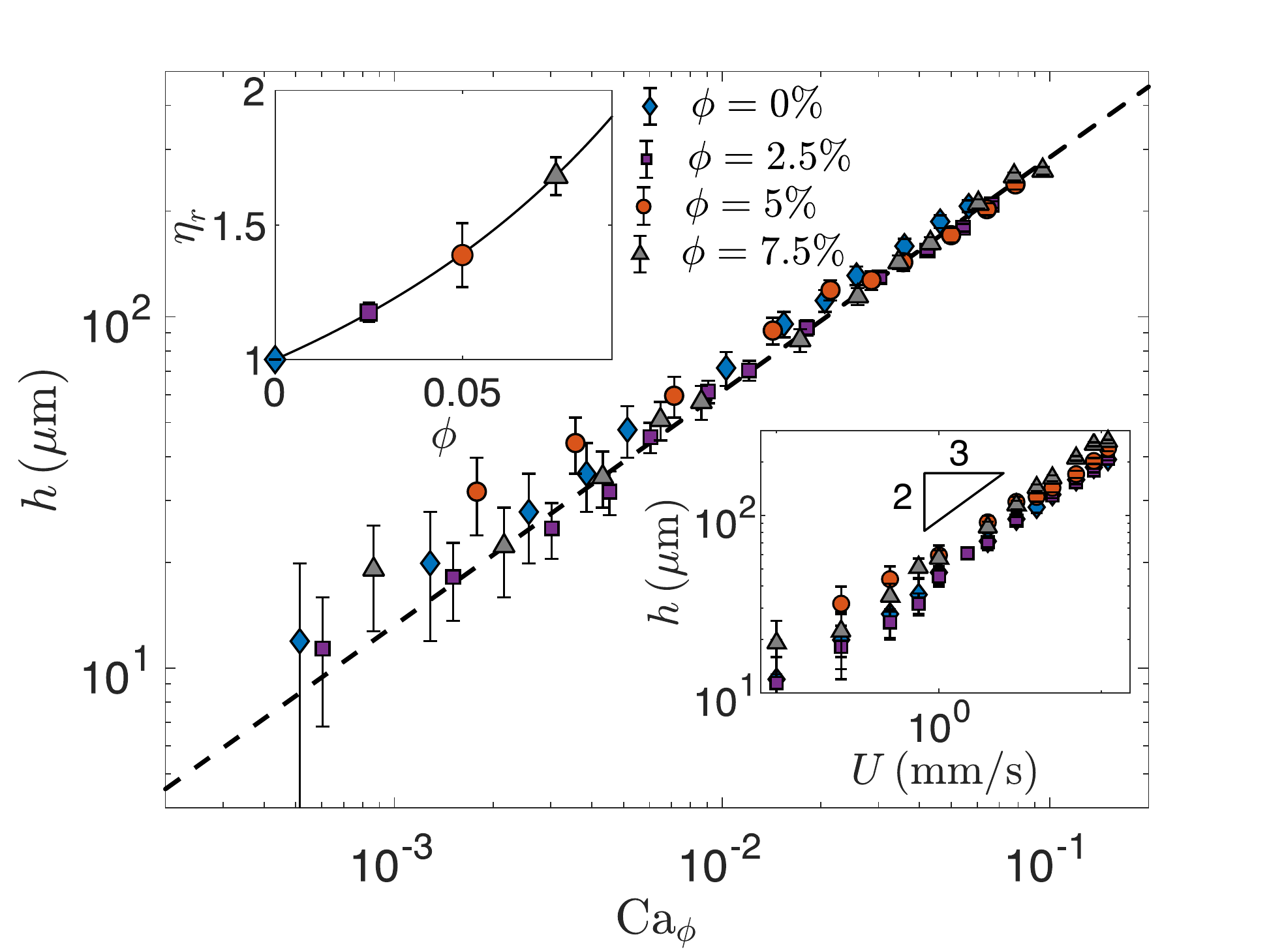}
  \caption{Coating film thickness $h$ on a flat substrate as a function of the effective capillary number $\mathrm{Ca}_{\phi}$ for various volume fraction $\phi$ (blue diamonds: $\phi=0\%$; purple squares: $\phi=2.5\%$; orange circles: $\phi=5\%$; grey diamonds: $\phi=7.5\%$). The dashed line represents the LLD law [Eq. (\ref{eq:LLD})]. The relative viscosity $\eta_r$ of the suspension is fitted to collapse the experimental data to the LLD law. Bottom inset: Thickness of the coating film $h$ as a function of the withdrawal velocity $U$. Top inset: Evolution of the relative viscosity $\eta_{\rm r}$ of fiber suspensions when varying the volume fraction $\phi$. The solid line represents the Maron-Pierce model [Eq. (\ref{eq:MP})] with $\phi_{\rm c} = 33\%$. The fibers used here have a diameter $d=50\,\mu\textrm{m}$ and a length $L=200\,\mu\textrm{m}$.}
  \label{fgr:LLD}
\end{figure}

As a result, we use an approach similar to the one used in Jeong \textit{et al.}\cite{jeong2022dip} More specifically, we make no assumption on the effective relative viscosity of the suspension, $\eta_{\rm r}(\phi)=\eta(\phi)/\eta_0$. Instead, $\eta_{\rm r}$ is treated as a fitting parameter in the LLD:
\begin{equation} \label{eq:LLD_New_phi}
h=0.94 \,\ell_{c}\, {\mathrm{Ca}_{\phi}}^{2 / 3}=0.94\, \ell_{c}\,\left(\frac{\eta(\phi)\, U}{\gamma}\right)^{2 / 3}.
\end{equation}
Here, the capillary number $\mathrm{Ca}_{\phi}=\eta(\phi)U/\gamma$ is based on the effective viscosity of the suspension and is different from $\mathrm{Ca}=\eta_0\,U/\gamma$ used in the previous sections. The two are related through $\mathrm{Ca}_{\phi}=\eta_{\rm r}(\phi)\,\mathrm{Ca}$. In addition, the presence of particles does not modify the surface tension, and the capillary length is thus unchanged.\cite{couturier2011suspensions,chateau2021extensional}

The main panel in figure \ref{fgr:LLD} reports the coating thickness for different volume fractions as a function of the effective capillary number $\mathrm{Ca}_{\phi}$. We fitted the values of $\eta_{\rm r}(\phi)$ so that the experimental measurements of the liquid thickness match the LLD prediction in the thick film regime (see also Jeong \textit{et al.}\cite{jeong2022dip}). As expected, the scaling ${\mathrm{Ca}_{\phi}}^{2/3}$ is recovered in the thick film regime since we previously noticed that $h \propto U^{2/3}$.

We report the values of the fitted relative viscosities for the different volume fractions in the top inset of figure \ref{fgr:LLD}. Similarly to the case of spherical particles, the relative viscosity increases with increasing fiber volume fractions $\phi$. The Maron-Pierce model, given by Eq. (\ref{eq:MP}), is used to fit the evolution of $\eta_{\rm r}$ and captures satisfactorily well the observed trend for $\eta_{\rm r}(\phi)$ when using a value of $\phi_c \sim 33\%$. As expected, the value of $\phi_c$ found here is smaller than for spherical particles. This shows that, qualitatively, a similar trend as Bounoua \textit{et al.}\cite{bounoua2019shear} is captured in the thick film regime. We should note that, in our case, the value of $\phi_c$ obtained is slightly lower than the result with a similar aspect ratio of fibers measured by Bounoua \textit{et al.}\cite{bounoua2019shear} which could be due to the alignment of the fiber due to the geometry of the flow, and some possible shear thinning.

In summary, a thick film regime is observed for fiber suspensions when the thickness of the coating film becomes thicker than the diameter of the fiber, \textit{i.e.}, $h \gtrsim d$. In this regime, similar to the observations made for suspensions of spherical particles,\cite{gans2019dip,palma2019dip,jeong2022dip} the thickness of the coating film can be predicted through the LLD law [Eq. (\ref{eq:LLD_New_phi})], where one needs to introduce an effective capillary number $\mathrm{Ca}_{\phi}=\eta(\phi)U/\gamma$ based on the viscosity of the fiber suspension.


\section{Conclusions} \label{sec:conclusion}

In this paper, we characterized the dip coating process of suspensions of non-Brownian fibers on flat and curved substrates. We first considered dilute suspensions to minimize the interaction between fibers. We considered the conditions for entrainment and the resulting orientation of a fiber in the coating film. We then investigated more concentrated fiber suspensions to characterize the thickness of the coating film.

Our experiments revealed that the entrainment of fibers on the surface of a substrate by dip coating is primarily controlled by the diameter of the fibers on both planar and cylindrical substrates. Similarly to previous results with suspensions of spherical particles,\cite{sauret2019capillary} we demonstrated that a fiber is entrained in the coating film if the thickness at the stagnation point is of order of the fiber radius $d/2$. For a flat substrate, this led to a condition on the capillary number ${\rm Ca} \geq {\rm Ca}^* =\beta \,[d/(2\,\ell_{\rm c})]^2$. The entrainment threshold increases slightly with the length $L$ of the fiber, but in a less significant manner. As a result, to predict the entrainment of fibers during the dip coating process, one could base their estimate on the diameter of the fiber only. We then reported that a similar argument remains valid to predict the entrainment of fibers on cylindrical rods. However, in this case, when the length of the fiber $L$ becomes larger than the radius of the rod $R$, the fibers need to be aligned along the axis of the rod, \textit{i.e.}, perpendicular to the meniscus, to be entrained. As a result, the entrainment threshold $U^*$ is higher for large values of $L/R$.

Compared to spherical particles, a particular feature of fibers is their anisotropy. We investigated the angular orientation of the fibers coating the substrate after the withdrawal process. On a flat substrate, deposited fibers do not show a preferential alignment regardless of the length of the fiber length when the thickness at the stagnation point $h^*$ is larger than the diameter of the fiber, which corresponds to $h \gtrsim d/3$. However, when the film becomes thinner and $h^*<d$, the fiber reorients when arriving at the meniscus due to capillary effects. One end of the fiber remains pinned, and large angles $\theta$ corresponding to fiber aligned with the meniscus are less commonly observed.

On cylindrical rods, the alignment of entrained fibers depends on the ratio of the fiber length $L$ to the radius of curvature of the substrate $R$. Whereas for $L/R \lesssim 1$, the situation observed is quite similar to the one reported for flat substrate, a preferential alignment is observed for $L/R \gtrsim 1$ (\textit{i.e.}, longer fibers and smaller substrate radius of curvature). For $L/R \gtrsim 1$, the fibers tend to align along the withdrawal direction. This observation is due to the fact that the fibers considered here are rigid and cannot wrap around the substrate.

We have finally considered more concentrated fiber suspensions and reported the presence of a thick film regime comparable to the one reported for spherical particles.\cite{gans2019dip,palma2019dip} In this regime, one can still use the LLD law by introducing an effective capillary number based on the viscosity of the fiber suspension.

In summary, the results presented here with fiber suspensions share many common features with the observations reported for spherical particles.\cite{gans2019dip,palma2019dip,jeong2022dip} The length scale to consider at first order is the diameter of the fibers $d$. The length $L$ appears to have an effect on the alignment on cylindrical substrates. The results presented here provide guidelines for manufacturing processes and show the conditions to control passively the orientation of fibers coating a substrate during the dip coating process.

\section*{Acknowledgements}
This material is based upon work supported by the National Science Foundation under NSF CAREER Program Award CBET Grant No. 1944844.


\balance

\bibliography{biblio} 
\bibliographystyle{unsrt}

\end{document}